\documentclass[aps,prx,twocolumn,superscriptaddress,showpacs,floatfix]{revtex4-1}
\usepackage{color}
\usepackage{mathtools}
\usepackage[caption=false]{subfig}
\usepackage{bm}        \usepackage{amssymb}   \usepackage{amsmath}
\usepackage[colorlinks=true,citecolor=blue,linkcolor=magenta]{hyperref}
\usepackage{graphicx}
    \usepackage[caption=false]{subfig}
\usepackage{braket}
\usepackage{natbib}
\usepackage [latin1]{inputenc}
\usepackage{siunitx}
\usepackage{blkarray, bigstrut}
\usepackage{xcolor}

\newcommand{\bea}{\begin{eqnarray}}
\newcommand{\eea}{\end{eqnarray}}

\newcommand{\be}{\begin{equation}}
\newcommand{\ee}{\end{equation}}

\newcommand{\beal}{\begin{align}}
\newcommand{\eeal}{\end{align}}
\newcommand{\ra}{\rangle}

\newcommand{\upa}{\uparrow}
\newcommand{\dna}{\downarrow}
\newcommand{\vS}{\vec{S}}
\newcommand{\dg}{{\dagger}}
\newcommand{\pdg}{{\phantom\dagger}}

\newcommand{\eq}[1]{Eq.~\ref{#1}}
\newcommand{\Sec}[1]{Sec.~\ref{#1}}
\newcommand{\app}[1]{App.~\ref{#1}}

\newcommand{\fig}[1]{Fig.~\ref{#1}}
\newcommand{\Fig}[1]{Fig.~\ref{#1}}
\def\Id{\mathbf{1}}
\newcommand{\transpose}{\top}
\def\tr{\mathrm{Tr}}

\ifdefined\added
\else
\usepackage[normalem]{ulem}
\newcommand{\added}[2][]{\textcolor{blue}{#2}\textsuperscript{\small\textcolor{red}{#1}}}
\definecolor{shadecolor}{RGB}{80,100,80}
\definecolor{pink}{RGB}{220,100,100}
\usepackage{marginnote}
\newcounter{mparcnt}

\fi

\def\PHo{\mathcal{P}_0}
\def\PH{\mathcal{P}}
\def\USW{U_{\PH\to\PHo}}
\def\Heff{H_{\text{eff}}}
\def\Hspin{H_{\text{spin}}}

\def\S{\tilde{\mathbf{S}}}
\def\Sc{\tilde{S}}
\def\h{\mathbf{h}}

\renewcommand{\vec}[1]{\mathbf{#1}}
\def\vS{\vec{S}}

\begin{document}

\title{Octupolar order and  Ising quantum criticality tuned
by strain and dimensionality: Application to $d$-orbital
Mott insulators}
\author{Sreekar Voleti}
\email{svoleti@physics.utoronto.ca}
\affiliation{Department of Physics, University of Toronto, 60 St. George Street, Toronto, ON, M5S 1A7 Canada}
\author{Arijit Haldar}
\email{arijit.haldar@utoronto.ca}
\affiliation{Department of Physics, University of Toronto, 60 St. George Street, Toronto, ON, M5S 1A7 Canada}
\author{Arun Paramekanti}
\email{arunp@physics.utoronto.ca}
\affiliation{Department of Physics, University of Toronto, 60 St. George Street, Toronto, ON, M5S 1A7 Canada}
\date{\today}
\begin{abstract}
Recent experiments have discovered multipolar orders in a variety of $d$-orbital Mott insulators. 
Motivated by uncovering the exchange interactions which underlie octupolar
order proposed in the osmate double perovskites, we study a two-site model using exact 
diagonalization on a five-orbital Hamiltonian, incorporating spin-orbit coupling (SOC) and interactions, and including both
intra-orbital and inter-orbital hopping. Using an exact Schrieffer-Wolff transformation, we then extract
an effective pseudospin Hamiltonian for the 
non-Kramers doublets,
uncovering dominant ferrooctupolar coupling driven by the interplay 
of two distinct intra-orbital
hopping terms. Using classical Monte Carlo simulations on the face-centered cubic lattice, we obtain a ferrooctupolar
transition temperature which is in good agreement with experiments on the osmate double perovskites. We also
explore the impact of uniaxial strain and dimensional tuning via ultrathin films, 
which are shown to induce a transverse field on the Ising octupolar order. This
suppresses $T_c$ and potentially allows one to access octupolar
Ising quantum critical points. We discuss possible implications of our results for a broader class of materials which
may host such non-Kramers doublet ions.
\end{abstract}
\pacs{75.25.aj, 75.40.Gb, 75.70.Tj}
\maketitle
\section{Introduction}
Multipolar orders have been extensively studied in $f$-electron compounds
\cite{MultipolarRMP2009,HauleKotliar2009,SantiniNpO2_PRL2000,NpO2TripleQ_PRL2002,Fazekas_PRB2003, NpO2NMR_PRL2006,Arima2013,Sakai_JPSJ2011,Sato_PRB2012,Nakatsuji_PRL2014,Blumberg_Science2015,Blumberg_PRL2016,Hattori2016,Freyer2018,SBLee2018,Patri2019}
where spin-orbit coupling and interactions dominate over weaker crystal field effects. However, there is growing evidence for
such exotic ``higher multipoles'' in a wide range of heavy $d$-orbital metals
such as LiOsO$_3$ and Cd$_2$Re$_2$O$_7$ which may exhibit
odd-parity nematic orders \cite{LFu_PRL2015,Hsieh_Science2017},
or quadrupolar orders as proposed
in A$_2$OsO$_4$ (with A = K,Rb,Cs) \cite{Motome2018}.

Recent work from various groups have also begun to explore such orders in the Mott insulator regime, where a local picture provides
a useful starting point.
For $d$-orbitals in an octahedral crystal field, the $t_{2g}$ single particle levels are split by SOC, resulting in 
a four-fold degenerate, $j_{\rm eff}  \!=\! 3/2$, ground state 
and a 
doubly degenerate, $j_{\rm eff}  \!=\! 1/2$, excited state. These levels can realize interesting multipolar phases at 
different electron fillings.
For instance, $d^1$ Mott insulators can realize the magnetism of
$j_{\rm eff}=3/2$ spins. Theoretical studies of such moments 
on the FCC lattice have shown that they can lead to 
wide regimes of quadrupolar order \cite{ChenBalents2010,ChenBalents2011,svoboda2021}
which may coexist with conventional
dipolar magnetic order, or valence bond orders \cite{RomhanyiPRL2017}. 
Experiments on $5d^1$ oxides, Ba$_2$NaOsO$_6$ with 
Os$^{7+}$ \cite{Mitrovic_NComm2017,Mitrovic_Physica2018} 
and Ba$_2$MgReO$_6$ with Re$^{6+}$ \cite{Hiroi_JPSJ2019}, have 
found evidence for two phase transitions, with a higher 
temperature quadrupolar ordering transition followed by 
dipolar ordering at a lower temperature.

In this paper, we focus on $d^2$ Mott insulators, where \cite{ChenBalents2010,ChenBalents2011,Svoboda_PRB2017,Lovesey2021} have argued for a local $J\!=\!2$ spin moment, which can lead to
various exotic orders including quadrupolar phases.
We have recently reexamined this issue 
\cite{paramekanti2019octupolar,voleti2020}
and shown that virtual excitations 
into the high
energy $e_g$ orbitals split
the five-fold degeneracy of the
$J\!=\! 2$ moment as
$2 (E_g) \oplus 3 (T_{2g})$, resulting in a ground state
non-Kramers $E_g$ doublet carrying quadrupolar and 
octupolar moments.
We had proposed, on phenomenological grounds, 
that ferro-octupolar (FO) order of these 
local moments
provides a comprehensive understanding \cite{paramekanti2019octupolar,voleti2020} of the
time-reversal breaking phase transition observed in 
the cubic ordered
double perovskite (DP) Mott insulators, Ba$_2$ZnOsO$_6$, Ba$_2$CaOsO$_6$, and Ba$_2$MgOsO$_6$, which host a $5d^2$ configuration on Os
\cite{Thompson_JPCM2014,Kermarrec2015,Thompson_PRB2016,MarjerrisonPRB2016,maharaj2019octupolar}. It is tempting to speculate that 
this Ising ferro-octupolar order might provide a template for 
storing
information. Interestingly, our theory of
octupolar order is reminiscent of, but distinct from, 
an old proposal by van den Brink 
and Khomskii \cite{khomskii2001} of ``complex $e_g$ orbital'' order in the colossal magnetoresistive manganites, which explored
time-reversal breaking in the single-particle 
$e_g$ orbitals.

Despite the seeming success of our proposal, 
our previous work did not fully
identify the microscopic origin of 
the octupolar exchange, although it did correctly identify
the mechanism by which quadrupolar exchange can get suppressed. 
In particular, there was no theoretical 
basis starting from a model of interacting spin-orbit
coupled electrons. This gap has been partially filled
by a very recent study which combines
density functional theory (DFT) 
and dynamical mean field theory (DMFT) calculations \cite{pourovskii2021}, and finds 
unequivocal evidence of FO exchange - however,
it is still desirable to clarify the origin of FO order
using a model tight-binding Hamiltonian.
Meanwhile,
several competing theories have emerged for the phase
transition observed in these osmates. One proposal
argues for antiferro-octupolar ordering of the $E_g$ doublets \cite{Lovesey2020}.
Other studies have argued for
antiferro-quadrupolar orders
based on a second-order perturbation theory 
calculation of the exchange interactions between non-Kramers 
doublets,
while also including coupling to Jahn-Teller active phonons \cite{Kee2021,churchill2021quadrupolar}.
However, the latter results do not naturally explain
the time-reversal
symmetry breaking observed in experiments. 
Motivated by these developments, we
consider here a five-orbital model for two neighboring sites,
which we solve using numerical
exact diagonalization (ED) to extract the exchange 
interactions. We find dominant FO 
exchange, in
qualitative agreement with the DFT and DMFT study \cite{pourovskii2021}, and identify a combination of
two distinct intra-orbital hoppings as the driving force
for FO exchange. Our results for the weaker
quadrupolar terms do not precisely match the DFT and DMFT
study \cite{pourovskii2021}; we attribute these differences
to differences in methodology. However, our ED results
are strikingly
different from the simple second-order perturbation 
projected to the $E_g$ doublets which finds dominant
quadrupolar exchange
\cite{Kee2021,churchill2021quadrupolar}. We show that this
discrepancy arises from the strong influence of the 
energetically close $T_{2g}$ triplets, which necessitates 
including higher order terms.
Armed with our ED results, we use Monte Carlo (MC) simulations
to explore the phase diagram as
we vary the
inter-orbital and intra-orbital hoppings. Over a wide 
regime of parameters, we find robust ferro-octupolar
order with high $T_c$, thus providing an explanation
for experimental observations on the double perovskite 
osmates. We also investigate the impact of uniaxial
strain and dimensionality, showing that this leads to
a transverse field on the Ising octupolar order,
allowing one to tune $T_c$ and potentially access
octupolar Ising quantum critical points. Our study extends
previous work showing strain-tuning of nematic (quadrupolar)
order and its transverse field quantum criticality \cite{Maharaj2017}.

This paper is organized as follows. In Section \ref{sec:pseudoHam} we discuss the single-site and two-site exact diagonalization results for the 
full five-orbital model, and show how we extract the pseudospin exchange model using an exact Schrieffer-Wolff transformation. 
Our results
yield large swaths of parameter space with dominant
FO exchange interactions on the face-centered cubic (FCC) 
lattice. In Section \ref{mc}, we discuss MC simulations of this pseudospin model, and show that it leads to a phase 
transition into the
FO ordered state with $T_c$ in reasonable agreement with
experiments on Ba$_2$ZnOsO$_6$, Ba$_2$CaOsO$_6$, and Ba$_2$MgOsO$_6$. Section \ref{sec:strain} studies the impact of uniaxial
strain and dimensional tuning via thin films, 
showing that it leads to an effective transverse field 
on the Ising FO order, suppressing $T_c$ and driving the system
towards an Ising quantum critical point. Section \ref{sec:discuss} presents
the summary and outlook.

\section{Pseudospin Hamiltonian}\label{sec:pseudoHam}

\subsection{Single-site exact diagonalization study} \label{singlesite}
The single-site model for the $d^2$ configuration incorporating both crystal-field effects, electron-electron interactions, and 
spin-orbit coupling, has been carefully explored in our
previous work. To keep our discussion self-contained, we sketch
the main results. We employ a single-site (local) Hamiltonian:
\bea
H_{\rm loc}=H_{\rm CEF}+H_{\rm SOC}+H_{\rm int}
\label{eq:hsingle}
\eea
which includes $t_{2g}-e_g$
crystal field splitting, SOC, and electronic interactions, written in the orbital basis 
($\{yz,xz,xy\},\{x^{2}\!-\!y^{2},3z^{2}\!-\!r^{2}\} ) \leftrightarrow (\{1,2,3\},\{4,5\}$) where $\alpha\equiv \{1,2,3\}$ label $t_{2g}$
orbitals and $\alpha\equiv \{4,5\}$ label $e_g$ orbitals.
The CEF term is given by:
\begin{equation} \label{cfham}
H_{\rm CEF}=V_C\sum_{\alpha=4,5}\sum_{s}n_{\alpha,s}
\end{equation}
where $s$ is the spin. The SOC term is
\begin{align}\label{eq:Hint}
\begin{split}
H_{\rm SOC} &= {\frac{\lambda}{2}} \sum_{\alpha, \beta} \sum_{s,s'} \bra{\alpha}\mathbf{L}\ket{\beta} \cdot \bra{s}\pmb{\sigma}\ket{s'}c^\dagger_{\alpha s} c_{\beta s'} \ ,
\end{split}
\end{align}
where $\pmb{\sigma}$ refers to the vector of Pauli matrices, and $\mathbf{L}$ are orbital angular momentum matrices. The operators $c_{\alpha s}$, $c^\dagger_{\alpha s}$ and $n_{\alpha s}$ destroy, create, and count the electrons with spin $s$ in orbital $\alpha$. The Kanamori interaction is given by
\bea
H_{\rm int} &=& U\sum_{\alpha}n_{\alpha \uparrow}n_{\alpha \downarrow} \!+\! \left( U' - {J_H \over 2} \right)  \sum_{\alpha > \beta} n_\alpha n_\beta 
 \\
 \!&-&\! J_H \sum_{\alpha \neq \beta} \vS_\alpha \cdot \vS_\beta \nonumber 
+ J_H \sum_{\alpha\neq\beta} c^\dg_{\alpha \upa} c^\dg_{\alpha\dna} c^\pdg_{\beta \downarrow} c^\pdg_{\beta \upa}
\eea
where $U$ and $U'$ are the intra- and inter-orbital Hubbard interactions, $J_H$ is the Hund's coupling, and $\vS_\alpha = (1/2) c^\dg_{\alpha s} \pmb{\sigma}_{s,s'} c^\pdg_{\alpha s'}$. The operator $n_\alpha\equiv n_{\alpha\uparrow}+n_{\alpha\downarrow}$ counts the total number of electrons in orbital $\alpha$. Assuming spherical symmetry of the 
Coulomb interaction, we set $U' = U - 2 J_H$ \cite{Georges2013}. In this calculation, we use $V_C = 2.2$ eV, $\lambda = 0.4$ eV, $U=2.5$ eV, and $J_H=0.3$ eV in order to obtain a spin gap (described below) which matches values obtained by neutron studies \cite{maharaj2019octupolar}.  

When the crystal field splitting $V_C \to \infty$, it leads to a five-fold degenerate ground state corresponding to a 
spin-orbit coupled $J=2$ 
quantum spin. For realistic finite $V_C$, 
this $J=2$ manifold is split,
leading to a
non-Kramers pseudospin doublet, with wavefunctions given in terms of $J_z$ eigenstates
as:
\bea\label{eq:nkramers_doublet}
|\psi_{g,\uparrow}\rangle = |0\rangle;~~~
|\psi_{g,\downarrow}\rangle = \frac{1}{\sqrt{2}} (|2\rangle + | -2 \rangle)
\eea 
and an excited state triplet separated from the doublet by a gap $\sim 20$\,meV.
The states $|\psi_{g,\upa}\ra,|\psi_{g,\dna}\ra$ are
individually time-reversal invariant.
The angular momentum operators $(J_x^2-J_y^2)/2\sqrt{3}$ and $-(3 J_z^2-J^2)/6$, restricted to this basis, act as
Pauli matrices $(\tau_x, \tau_z)$, forming the two
components of an XY-like quadrupolar order parameter, while $-\overline{J_x J_y J_z}/3$ (with overline denoting symmetrization) behaves
as $\tau_y$, and serves as the Ising-like octupolar order parameter.
We will define the corresponding pseudospin-$1/2$ operators as
$\tilde{S}_\alpha = \tau_\alpha/2$.
The ferro-octupolar order discussed later corresponds to all pseudospins being in the state 
$|\psi^{\rm oct}_\pm\ra = |\psi_{g,\upa}\ra \pm i |\psi_{g,\dna}\ra$, with the signs reflecting the $\mathbb{Z}_2$ Ising character of octupolar order, and the factor of `$i$' 
reflecting time-reversal symmetry breaking. 

Our next goal is to uncover the interaction between these
pseudospins on neighboring sites.

\subsection{Two-site exact diagonalization calculation}

\begin{figure}[!h]
    \centering
\includegraphics[width=0.4\textwidth]{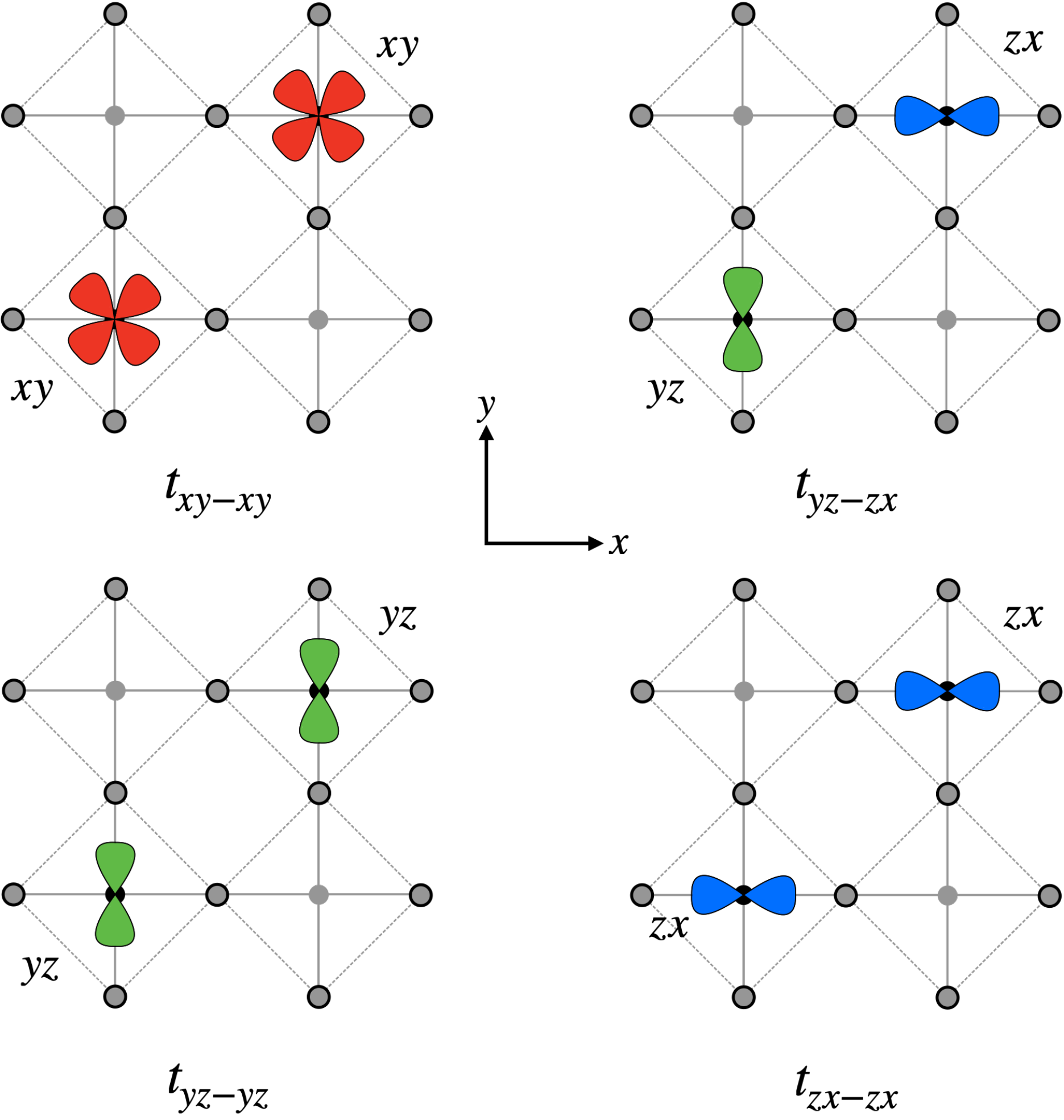}
    \caption{Schematic top-view of orbitals in the $xy$ plane showing different nearest-neighbor hoppings in the $xy$ plane. The corresponding hoppings in the $xz, yz$ planes are determined 
    by the cubic point group symmetry.}
    \label{fig:hopping}
\end{figure}

We consider a two-site model, with each site housing a non-Kramers doublet as described above. Two sites lying in the $\gamma$ plane (where $\gamma \in \{ xy , yz , zx\}$) are coupled via a hopping Hamiltonian of the form

\bea
\label{hopham}
H_T^\gamma = \sum_{\alpha \beta s}(T^{\gamma}_{\alpha \beta} c^\dg_{2\beta s} c^\pdg_{1\alpha s} + T^{\gamma \dagger}_{\beta \alpha} c^\dg_{1\alpha s} c^\pdg_{2\beta s })
\eea

where $T^{\gamma}$ is the hopping matrix in the $\gamma$ plane. In the $xy$ plane the sites are coupled via four hopping channels, as described in \fig{fig:hopping}. The matrix in this plane takes the form 

\newcommand{\txyxy}{t_{xy\mbox{-}xy}}
\newcommand{\tzxzx}{t_{zx\mbox{-}zx}}
\newcommand{\tyzyz}{t_{yz\mbox{-}yz}}
\newcommand{\tyzzx}{t_{yz\mbox{-}zx}}
\newcommand{\tzxyz}{t_{zx\mbox{-}yz}}

\bea 
\label{eq:hopmat}
T^{xy} = 
\left( \begin{array}{ccc|cc}
\tyzyz & \tyzzx & 0 & 0 & 0 \\
\tzxyz & \tzxzx & 0 & 0 & 0 \\
0 & 0 & \txyxy & 0 & 0 \\
\hline 
0 & 0 & 0 & 0 & 0 \\
0 & 0 & 0 & 0 & 0 
\end{array} 
\right)\ .
\eea
With cubic symmetry, the corresponding matrices in the other planes can be obtained via $C_3$ transformations about the [111] direction.
The dominant hopping in the $xy$ plane is
$\txyxy$, which is larger than $\tyzzx=\tzxyz$, which in
turn is larger than $\tyzyz=\tzxzx$.
The two-site model can be studied in the Fock 
space sector with four $d$ electrons having 
access to 20 distinct states (10 on each site). 
The dimension of the resulting Hilbert space is thus $ {20 \choose 4} = 4845$. The two-site Hamiltonian 
\bea
\label{eq:hfull}
H^\gamma = H_{\rm loc} \otimes \mathcal{I} +  \mathcal{I} \otimes H_{\rm loc} + H_T^\gamma
\eea
consists of two copies of the single-site Hamiltonian presented in \eq{eq:hsingle}, in addition to $H_T^\gamma$. The symbol $\mathcal{I}$ denotes the identity operator for the Hilbert space in the single-site problem.

\subsection{Exact Schrieffer-Wolff transformation to obtain the pseudospin Hamiltonian}\label{sec:SW}

We compute the effective pseudospin Hamiltonian in the presence of
intersite couplings using an exact Schrieffer-Wolff (SW) transformation \cite{schriefferPRev1966, bravyiSW2011}.
SW transformations in general are used to obtain effective low-energy description
of a ``perturbed" Hamiltonian in terms of the low-energy eigenstates of the original
``unperturbed" Hamiltonian. This is accomplished by defining a so called
direct rotation \cite{bravyiSW2011} that connects the low-energy subspaces of the ``unperturbed" and
``perturbed" Hamiltonians.

For our two-site model,  the intersite hoppings in \eq{hopham} will serve as the source of the 
perturbation. Therefore, we consider the two-site decoupled Hamiltonian (\eq{eq:hfull} without $H_T^\gamma$ ) as the
unperturbed Hamiltonian $H_0$ and the coupled model $H^\gamma$ (\eq{eq:hfull}) as the perturbed Hamiltonian. As already discussed (\Sec{singlesite}), the low-energy subspace of the single-site Hamiltonian
has a two-fold degeneracy, which translates to a four-dimensional degenerate subspace for the
decoupled two-site Hamiltonian $H_0$. We refer to this subspace for the decoupled (unperturbed)
Hamiltonian as $\PHo$. Upon introducing intersite couplings, the subspace $\PHo$ gets modified
perturbatively to a different four-dimensional subspace $\PH$. This new subspace $\PH$ is
by definition the low-energy eigenspace for the coupled two-site Hamiltonian $H^\gamma$.

The SW transformation that rotates the subspace $\PH$ to $\PHo$ is defined as the unitary
transformation
\begin{align}\label{eq:USW}
	\USW = \sqrt{(2 P_0-\Id)(2 P-\Id)},
\end{align}
where $P=\sum_{\phi\in\PH}\ket{\phi}\bra{\phi}$, $P_0=\sum_{\psi_0\in\PHo}|\psi\rangle\langle\psi|$ are the projection operators onto
the subspace $\PH$ and $\PHo$ respectively, and $\Id$ is the identity operator. The square root in \eq{eq:USW} is defined using a branch cut
on the complex plane such that $\sqrt{1}=1$. The states $\{\ket{\phi}\}$ denote a choice of basis spanning $\PH$ and $\{\ket{\psi}\}$ is a basis for $\PHo$. The operator $\USW$ by construction maps a state
 $\ket{\phi} \in \PH$ to a unique state $\ket{\psi} \in \PHo$, such that $\USW\ket{\phi}=\ket{\psi}$. Consequently,
 $\USW^\dagger$ does the opposite, i.e., $\USW^\dagger\ket{\psi} = \ket{\phi}$.
Furthermore, $\USW$ is guaranteed to be unique \emph{iff} $\USW^2$, i.e. $(2 P_0-\Id)(2 P-\Id)$, does not have any eigenvalues
that reside on the negative real axis of the complex plane. It has been shown \cite{bravyiSW2011} that
this is indeed the case when the corrections arising from the perturbation are sufficiently small compared with
the spectral-gap $\Delta$ separating the low-energy subspace $\PHo$ from the excited states
of the \emph{unperturbed} Hamiltonian $H_0$. For our two-site model, as discussed below, we have checked that the perturbative 
level-shifts are weak compared with the spectral-gap $\Delta$ separating the non-Kramer's doublet (\eq{eq:nkramers_doublet}) from rest of the spectrum, justifying a pseudospin-$1/2$ model of the
low energy two-site spectrum.

Usually a direct computation of $\USW$ (see \eq{eq:USW}) is extremely difficult in a many-body setting, since a
full computation of the perturbed subspace $\PH$, spanning all orders of perturbation, is hard due to the exponential complexity of the many-body problem. Therefore, a series expansion for $\USW$ in powers of perturbation strength is often used as an approximation. However, for our two-site
problem the dimension of the many-body Fock space is 4845 (see discussion above \eq{eq:hfull}), and well within reach of exact diagonalization (ED) techniques. This allows us to solve for the low energy subspace $\PH$ ($\PHo$) for the perturbed (unperturbed) Hamiltonian exactly, and obtain $\USW$ using \eq{eq:USW} to all orders of perturbation in intersite couplings.

We then use the computed SW transformation $\USW$ to obtain the effective low energy form
of the perturbed Hamiltonian $H^\gamma$ in the original subspace $\PHo$ of the unperturbed problem, as
follows
\begin{align}
	\Heff^\gamma =\left(P_0 \USW\right) H^\gamma \left(\USW^\dagger P_0\right).
\end{align}
Since we use the exact SW transformation $\USW$ to compute $\Heff^\gamma$, the resulting $4\times4$ Hamiltonian
is also exact in the sense that it has contributions from all orders of perturbation. By construction, the eigenvalues of $\Heff^\gamma$ are precisely equal to the lowest four eigenvalues of $H^\gamma$.

Having proposed the strategy to extract $\Heff^\gamma$, we need to compute the effective $4\times4$ Hamiltonian in a basis that will naturally allow us to interpret $\Heff^\gamma$ in the form of a valid pseudospin Hamiltonian. Therefore, we carry out the entire computation 
discussed above, using a basis spanning $\PHo$ in which the operators $(J_x^2-J_y^2)$, $-\overline{J_x J_y J_z}$,  $(3 J_z^2-J^2)$ on sites $i=1,2$, admit the Pauli matrix representations
$\tau_x\otimes \tau_0$, $\tau_y\otimes \tau_0$, $\tau_z\otimes \tau_0$ and $\tau_0\otimes\tau_x$, $\tau_0\otimes\tau_y$, $\tau_0\otimes\tau_z$, respectively, where $\tau_0$ is the $2\times 2 $ identity matrix. The steps that go into selecting such a basis for $\PHo$ are discussed in \app{sw_appendix}. The resulting pseudospin Hamiltonian of the $\gamma$ plane in this basis takes the general form:
\begin{align} \label{swspinham}
	\Hspin^\gamma = \S^\transpose_1 \mathcal{K}^\gamma  \S_2 + \h_1^\gamma \cdot \S_1 + \h_2^\gamma\cdot \S_2,
\end{align}
where the symbols $\S_{i=1,2}\equiv[\Sc_{ix}, \Sc_{iy}, \Sc_{iz}]$ represent the pseudospin operators for the two sites $i=1,2$. The effective ``spin-spin" interactions are encoded in the $3\times 3$ $\mathcal{K}^\gamma$ tensor and $\h_i^\gamma$ are effective time-reversal even ``Zeeman'' fields acting on the pseudospins. Both, the $\mathcal{K}^\gamma$ tensor and the components of the fields $\h_i^\gamma$, can be obtained from the exactly computed $\Heff^\gamma$ as follows
\begin{align}
\begin{split}
	\mathcal{K}^\gamma_{\alpha\beta} &= \tr~[\Heff^\gamma \left(\tau_\alpha\otimes \tau_\beta\right)]\notag\\
	(\h^\gamma_1)_\alpha &= \tr~[\Heff^\gamma \left(\tau_\alpha\otimes\tau_0\right)]/2\notag\\
	(\h^\gamma_2)_\alpha &= \tr~[\Heff^\gamma \left(\tau_0\otimes\tau_\alpha\right)]/2    
\end{split}
\ .
\end{align}
While the ``Zeeman'' fields appear to break the cubic symmetry of the lattice, they appear precisely because we consider one bond at a time (Fig. \ref{fig:hopping}, for example, shows only the bond in the $xy$ plane), a process which does not respect cubic symmetry. When summed over all the neighbours of the FCC lattice, the net field vanishes exactly, i.e.
$$\sum_{\gamma\in\{xy,yz,zx\}}\h_i^\gamma=0 \ ,$$
thus restoring the full symmetry.
In \Sec{sec:strain}, we will consider the impact of uniaxial strain
or crystal surfaces, which will give rise to situations where the net ``Zeeman'' field on the pseudospin does not vanish 
(which is to be expected, since the strain explicitly breaks 
cubic symmetry and intersite couplings can then lift the pseudospin degeneracy). 
\subsection{Exchange couplings} \label{twositeresults}

\begin{figure}[!t]
\includegraphics[width=0.5\textwidth]{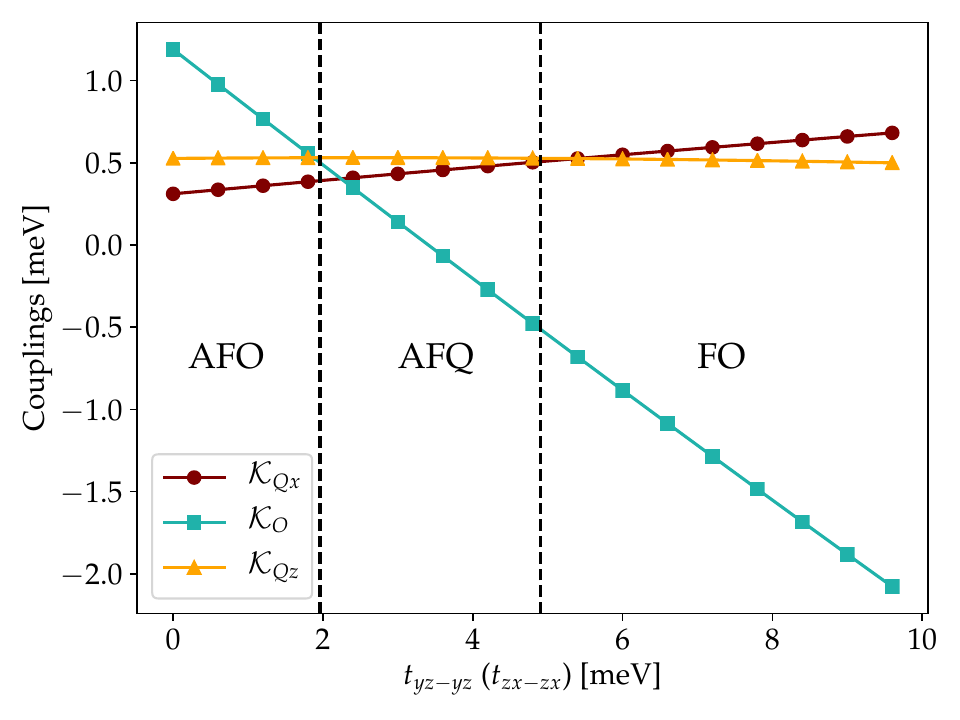}
    \caption{Evolution of couplings in $xy$ plane as a function of the intra-orbital $yz$-$yz$ and $zx$-$zx$ hopping. The dashed lines show the phases based on the most dominant couplings. This cut was made for $\txyxy\!=\!-150$ meV and $\tyzzx\!=\!30$ meV.}
    \label{fig:exchange}
\end{figure}

The symmetry considerations outlined in \app{symm_appendix} dictate that the pseudospin Hamiltonian in the $xy$ plane is an $XYZ$ model:
\bea\label{eq:Hspin}
H_{\rm spin}^{xy} = \mathcal{K}_{Qx} \Sc_{1x} \Sc_{2x} + \mathcal{K}_{Qz} \Sc_{1z} \Sc_{2z} + \mathcal{K}_O \Sc_{1y} \Sc_{2y}\ ,
\eea 
where $\mathcal{K}_{Qx}$ and $\mathcal{K}_{Qz}$ are the quadrupole-quadrupole couplings, and $\mathcal{K}_O$ is the octupole-octupole coupling. For bonds in other planes, the
exchange couplings may be obtained using $C_3$ rotations about
the (111) axis, and they involve off-diagonal symmetric
couplings of the form $(\Sc_{1x} \Sc_{2z} + \Sc_{1z} \Sc_{2x})$.
In this section, we will drop the Zeeman field terms from \eq{swspinham}, since they cancel out upon summing over all
neighbors as outlined at the end of \Sec{sec:SW}. Keeping the dominant $\txyxy$ hopping fixed at $-150$ meV, we vary $\tyzzx$ and $\tyzyz =\tzxzx$ in the ranges $0\!-\!30$ meV and $0\!-\!10$ meV, respectively, to study the dominant order
hosted by the pseudospin models.
We do so by analyzing the dependence of the couplings, $\mathcal{K}_{Qx}$, $\mathcal{K}_{Qz}$ and $\mathcal{K}_O$ on the hopping terms $\tyzyz$ and $\tyzzx$.
\Fig{fig:exchange} shows a representative example of this analysis when 
$\tyzzx=\ 30$ meV.
As a first pass at identifying the phases in the model, we simply assign phases based on the dominant coupling in the $XYZ$ model 
- an approach that will be corroborated below by classical 
MC simulations in \Sec{mc}. The three phases that appear in 
this phase diagram spanned by the subdominant hoppings are:
\begin{enumerate}
\item \!\! Ferro-Octupolar (FO): $\mathcal{K}_O \!<\! 0$
\item \!\! Antiferro-Octupolar (AFO): $\mathcal{K}_O \!>\! 0$
\item \!\! Antiferro-Quadrupolar (AFQ): 
$\!\mathcal{K}_{Q\alpha} \!\!> \!\! 0$ 
($\alpha\! \in\! \{x,z\}$).
\end{enumerate}

\begin{figure}[!t]
    \centering
\includegraphics[scale=0.99]{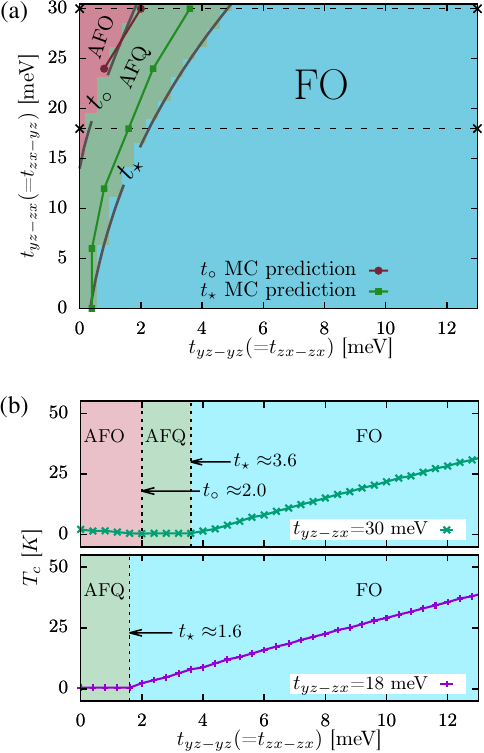}
    \caption{(a) Phase diagram in subdominant hopping parameter space, using a fixed $t_{xy-xy} = -150$ meV. The solid black phase boundaries $t_\circ$ and $t_\star$ denote the AFO-AFQ and AFQ-FO phase transition boundaries calculated simply by considering the dominant coupling (as outlined in \Sec{twositeresults}). The colored lines (represented with symbols -- {$\bullet$}, {\tiny$\blacksquare$}) denote the phase boundaries obtained from MC simulations (as outlined in \Sec{mc}). The MC predictions of the $t_\star$ boundary occur at smaller values of $t_{yz-zx}$, likely because the frustration of the quadrupoles on the FCC lattice causes the unfrustrated ferro-octupolar (FO) order to be preferentially stabilized despite the coupling having a smaller magnitude. (b) Representative cuts along the $y$-axis of the phase diagram (denoted as dashed-horizontal lines in (a)), showing the evolution of the critical temperature $T_c$ with $t_{yz-yz}$. Three piecewise regions, representing the indicated phases, can be observed. It can be seen that the values of $T_c$ for the AFO ($0<T_c<3 K$) and AFQ ($0<T_c<1 K$) phases are far lower than the $T_c$ values of 30-50$K$ reported in experiments, while the $T_c$ for the FO phase fits well with experiments.}
    \label{fig:pd}
\end{figure}

In \fig{fig:pd}(a), we lay out the full phase diagram in the $\tyzyz-\tyzzx$ plane and label the phases appropriately by identifying the dominant coupling.
Interestingly, we see that the FO phase forms the largest and most robust swath of the phase diagram centered around \fig{fig:pd}(a). Previous proposals have argued for the stabilization of both AFO and AFQ phases, and while these phase do exist in our model, we will show in \Sec{mc} that for the reasonable choice of hopping parameters used, they have critical temperatures that are incompatible with experimental evidence. The specific quadrupolar ordering patterns coming from these frustrated interactions have been explored in previous works \cite{Lovesey2020,Kee2021}.

\subsection{Comparison of exact and 
second-order perturbation theory results}
\begin{figure}[!t]
\includegraphics[width=0.45\textwidth]{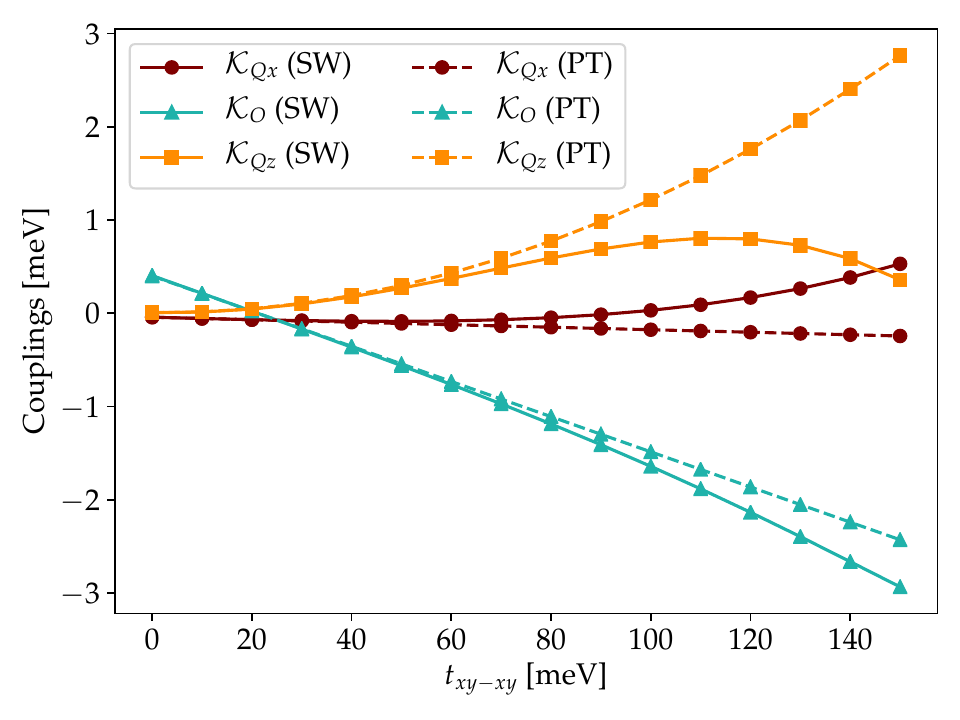}
    \caption{Comparison of the exchange couplings computed using the Schrieffer-Wolff (SW, solid lines) method and second order Perturbation Theory (PT, dashed lines). It can be seen that a second order treatment agrees with the SW results for small $t_{xy-xy}$, but would lead to the erroneous conclusion that the antiferromagnetic quadrupolar interactions would prevail, while in reality it gets suppressed and causes the ferro-octupolar interaction to dominate. We have fixed $t_{yz-zx}=18$\,meV and
    $t_{yz-yz}=t_{zx-zx}=10$\,meV.}
    \label{fig:exvsxyxy}
\end{figure}

One might reasonably wonder why it is necessary to use the ED
and Schrieffer-Wolff method to extract the exchange couplings
between the non-Kramers doublets, which goes beyond 
the standard second-order perturbation theory \cite{Kee2021}. 
In order to understand this, we note that while the charge gap is
indeed much larger than the hopping energy scale, there is a
small scale corresponding to the splitting between the 
non-Kramers ground state doublet and the excited triplet. Due to this small scale, hopping processes which involve an intermediate 
hopping to the 
triplet before returning to the doublet become significant. In perturbation theory, such processes occur at fourth order; a simple second order treatment completely misses these effects. As the energy splitting becomes smaller, it is conceivable that even higher order processes may become significant. To illustrate this point, we show in \fig{fig:exvsxyxy} the evolution of the coupling constants as a function of the dominant hopping $t_{xy-xy}$. It is clear that the
second order perturbation theory agrees with the exact calculation
for small $t_{xy-xy}$, but a further increase of $t_{xy-xy}$
leads to a suppression of the quadrupolar interactions which is
not captured by second order perturbation theory; this
suppression leads to the dominance of the 
ferro-octupolar exchange.

\section{Monte Carlo simulations on the face-centered cubic lattice} \label{mc}

In this section, we discuss the phase diagram of the
pseudospin-$1/2$ Hamiltonian in \eq{eq:Hspin}, with coupling constants derived from
microscopics,
by treating the pseudospins as
classical moments, and using MC simulations
to extract their ordering and thermal phase transitions. Such an
approach is expected to qualitatively capture the phase diagram
on the 3D face-centered cubic lattice of the ordered
double perovskites; quantum fluctuations may lead to
quantitative corrections to the phase boundaries and 
transition temperatures.

\begin{figure}[!t]
\includegraphics[width=0.45\textwidth]{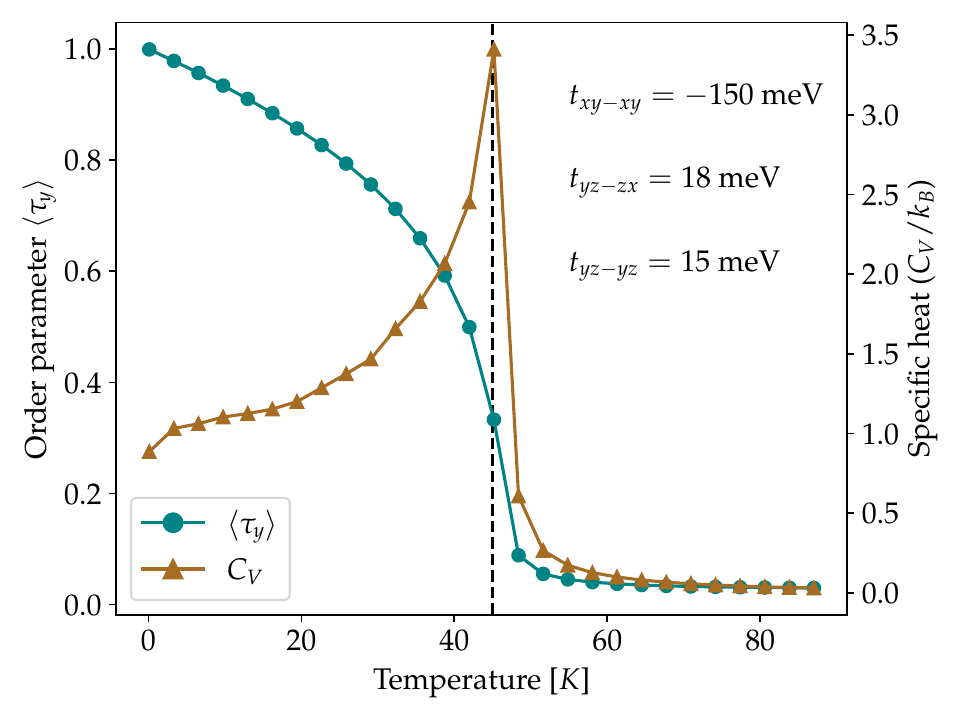}
    \caption{Representative example of MC simulation, for the hopping parameters corresponding to the Ferro-Octupolar (FO) phase. It can be seen that there is a single phase transition, marked by a peak in the specific heat ($C_V$) and the  emergence of a ferro-octupolar order parameter $\langle \tau_y 
    \rangle$.}
    \label{fig:mc}
\end{figure}

The simulations were conducted using the SpinMC package \cite{SpinMCBuessen} on a cluster of 1331 spins ($11 \times 11 \times 11$ primitive FCC cluster) with periodic boundary conditions, over a temperature range of $0.01$ to $10$ meV (corresponding to $0.116$ to $116$ K). To construct a phase diagram, pseudospin Hamiltonians were generated using a fixed dominant hopping $t_{xy-xy} = -150 $ meV, and varying $t_{yz-zx}$ and $t_{yz-yz}$ in the ranges $0\!-\!30$ meV and $0\!-\!20$ meV, respectively (DFT studies on these and other $5d$ double perovskites have shown that these are reasonable choices \cite{revelli2019,churchill2021quadrupolar}). We observe, in each case, a single thermal phase 
transition marked by a sharp peak in the specific heat $C_V$, and accompanied by the development of a nonzero 
order parameter as illustrated for a ferro-octupolar transition in \fig{fig:mc}. This representative plot was generated using hopping parameters close to those recently obtained using 
{\it ab initio} electronic structure calculations \cite{churchill2021quadrupolar} on the osmate double
perovskites.

To identify phase boundaries shown in the full phase diagram in \fig{fig:pd}(a), we look at the development of the critical temperature $T_c$ as a function of the hoppings $t_{yz-zx}$ and $t_{yz-yz}$ (see \fig{fig:pd}(b)), and identify kinks, suggesting a change in the underlying analytic form of the dependence. The phase boundaries (shown with solid-lines and symbols in \fig{fig:pd}(a)) obtained using this method match well with the ``na\"ive" method of computing the phase boundaries ($t_\circ$, $t_\star$ in \fig{fig:pd}(a)) described in \Sec{twositeresults}, where we simply looked at the dominant term in the pseudospin Hamiltonian. 

\section{Tuning octupolar order via uniaxial strain or dimensionality}\label{sec:strain}
The multipolar orders we have obtained above are highly sensitive to the nature of the inter-orbital and intra-orbital hoppings as discussed above. In addition, we have seen that the full cubic point group symmetry leads to a cancellation of the time-reversal even ``field'' terms acting on the $(\tau_x,\tau_z)$ pseudospin components, leaving us with only two-spin exchange terms. Motivated by tuning the multipolar orders, we next consider the impact of breaking cubic symmetry via strain or interfaces on the pseudospin Hamiltonian. 

\subsection{Uniaxial strain} 
Let us consider uniaxial strain along the (001) axis ($z$-axis), 
which we take into account by rescaling all the inter-site hoppings for neighbors in the $xz$ and $yz$ planes by a factor $(1-\delta)$, with $\delta > 0$ corresponding to tensile strain and $\delta < 0$ corresponding to compressive strain. Given the typical strong dependence of the hopping amplitudes on the lattice constants \cite{Grosso1995,WangNPJ2017}, we expect the lattice strain 
$\varepsilon_{zz} \ll \delta$. A careful account of strain effects must rely on experiments and {\it ab initio} electronic structure calculations, in order to relate $\delta$ to changes in lattice constants, and to examine changes in the relative strengths of the inter-orbital and intra-orbital terms; we defer this to a future study.
We repeat
the two-site exact diagonalization and 
Schrieffer-Wolff procedure as a function of $\delta$, and 
find that $\delta\neq 0$
leads to a non-cancelling ``field'' acting 
on the $(\tau_x,\tau_z)$ 
pseudospin components due to loss of cubic symmetry. 
This ``field'' is transverse to the octupolar ordering direction $\tau_y$, and can thus induce quantum fluctuations which can suppress $\langle \tau_y \rangle$, and potentially 
reveal a three-dimensional (3D) octupolar Ising quantum critical point. 

\begin{figure}[!t]
\includegraphics[width=0.45\textwidth]{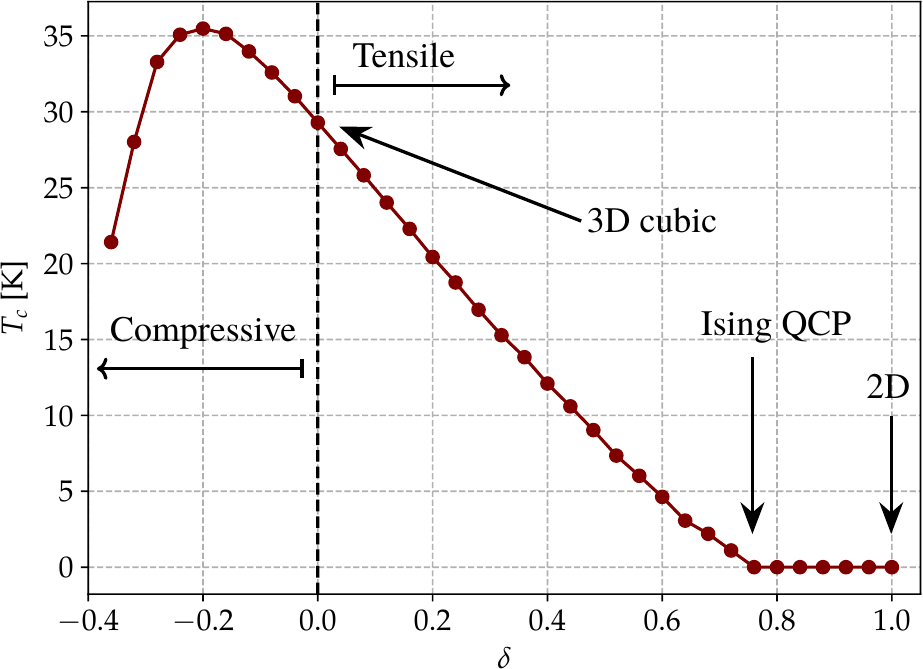}
    \caption{Evolution of the ferro-octupolar transition temperature, $T_c$, as a function of the hopping distortion $\delta$ induced by uniaxial strain. The cubic hopping parameters used here are $t_{xy-xy}=-150$\,meV, $t_{yz-zx}=18$\,meV, and 
    $t_{yz-yz}=10$\,meV. }
    \label{fig:tcdist}
\end{figure}

\Fig{fig:tcdist} shows the $T_c$ computed using
MC simulations with 
the modified exchange couplings and induced ``Zeeman field'' 
in the presence of strain. We find that tensile strain ($\delta > 0$)
leads to a strong suppression of $T_c$, due to combined effect of the weakening of octupolar exchange interactions in the $yz, zx$ planes and the generated transverse field. On the other hand, for compressive strain ($\delta < 0$), $T_c$ first increases, since the enhancement of the octupolar exchange coupling is initially more significant than the generated transverse field, before it begins to drop. For tensile strain, we find that the (mean-field) 3D Ising quantum critical point is
at $\delta_c \approx 0.8$, beyond which the 
octupolar ordering is suppressed.
While this critical point $\delta_c$ may not be accessible in experiments, the $\delta$-dependence of $T_c$ at smaller 
strain may be more easily tested. We note again that the strain
$\varepsilon_{zz} \ll \delta$.

Previous work has shown that shear strain can act as a transverse field on Ising nematic order and drive a nematic quantum phase transition \cite{Maharaj2017}. Our work generalizes this idea to the case of octupolar order. Our work also goes beyond previous studies which have explored the interplay of weak magnetic field and strain for probes of octupolar ordering or octupolar susceptibility \cite{Patri2019,sorensen2021}.

\subsection{Ultrathin films} 
The generation of ``Zeeman fields'' when symmetry is lowered from cubic also happens naturally at surfaces
or interfaces. In particular, let us consider an ultrathin (001) epitaxial film where
the top and bottom faces experience a field $\propto \tau_z$ due to reduced symmetry. If this surface field is sufficiently strong, and the film thickness is sufficiently small, the transverse surface field can kill the octupolar order in the entire film.

In order to properly take into account the effective fields that may appear at the surfaces, it is important to ensure that all the possible interactions which break cubic symmetry on a single bond are taken into account. In the preceding sections, we had
not included \textit{intersite} Coulomb interactions; as we have
explicitly checked, their inclusion has a negligible impact on the
exchange couplings. However, we find that
these interactions have a large effect on the effective fields
$\mathbf{h}^\gamma$ in \eq{swspinham}; while the
Coulomb-induced terms cancel in the bulk when we add up 
the contributions
from the twelve nearest-neighbors, this cancellation does
not occur at surfaces. In what follows, we incorporate 
these residual inter-site
Coulomb interactions and extract the transverse fields
at the surface; the calculation of these Coulomb
matrix elements for the osmate double perovskites is described in \app{app:Uab}.

\begin{figure}[!t]
\includegraphics[width=0.45\textwidth]{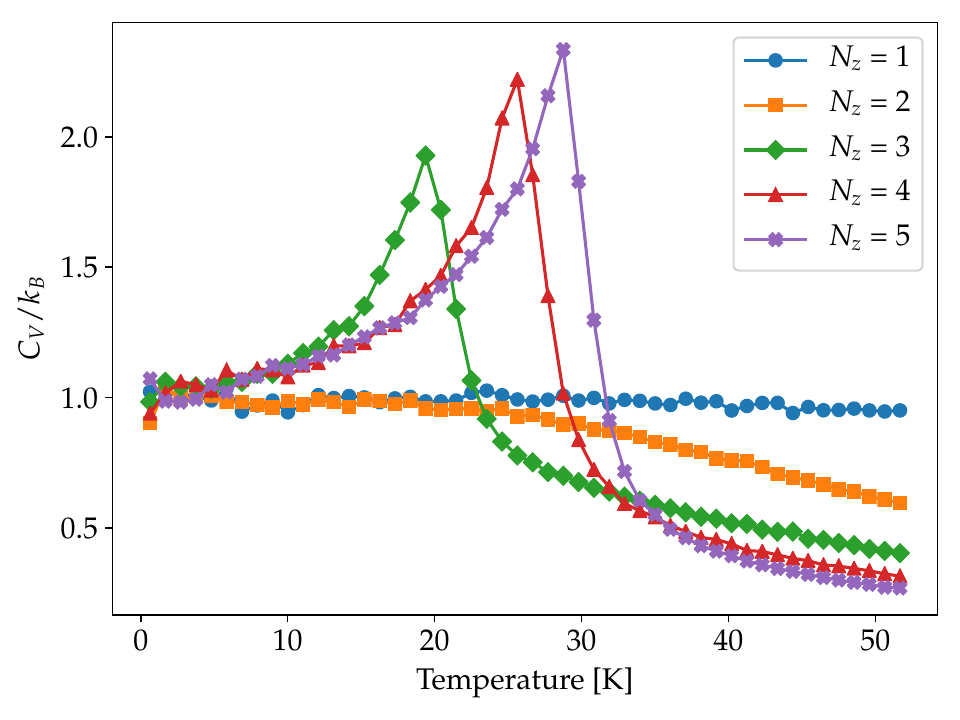}
    \caption{Evolution of the ferro-octupolar phase transition as a function of film thickness $N_z$.  The specific heat peak marking the transition indicates that $T_c$ decreases with film thickness, eventually vanishing for $N_z \leq 2$; see
    text for details.}
    \label{fig:cvevol}
\end{figure}

\begin{figure}[!t]
\includegraphics[width=0.45\textwidth]{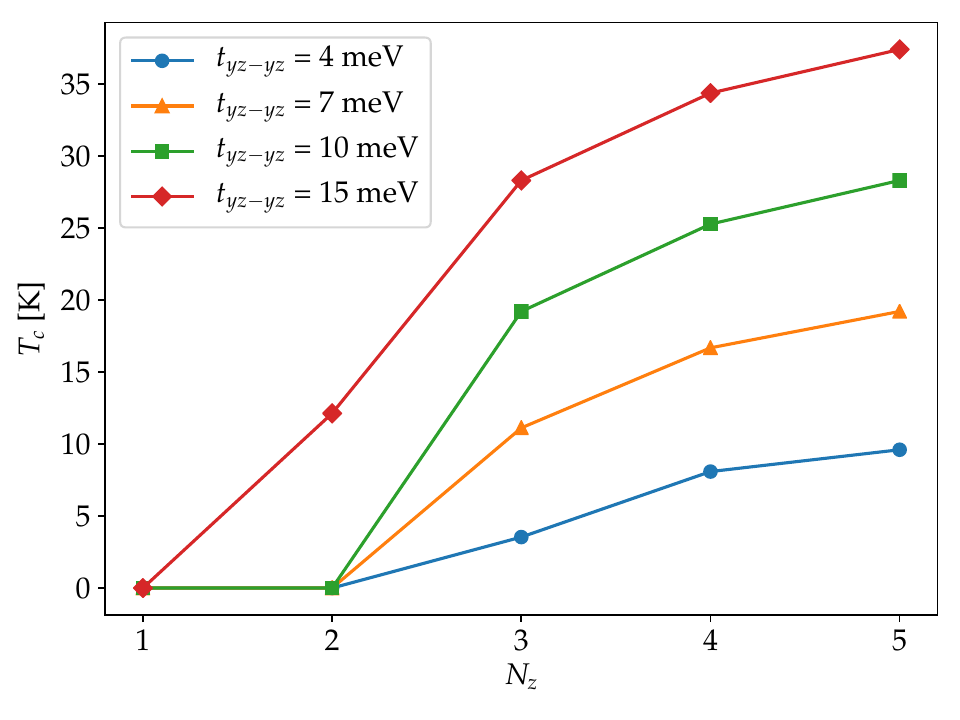}
    \caption{Evolution of the critical temperature, $T_c$, as a function of the film thickness, $N_z$, for various
    values of $\tyzyz$.}
    \label{fig:tcvsnz}
\end{figure}

With the above consideration in place, we consider a $12 \times 12 \times N_z$ lattice, with periodic boundary conditions 
along the $x$ and $y$ directions, and open boundary conditions along the $z$ direction. Here $N_z$ represents the film thickness. 
We incorporate the microscopically
computed effective transverse fields on
the top and bottom faces of the film.
For illustrative purposes, we fix two hopping parameters, $t_{xy-xy}=-150$ meV, $t_{yz-zx}=18$ meV, and vary $t_{yz-yz}$. \Fig{fig:cvevol} shows (for $t_{yz-yz}=10$ meV) the evolution of the specific heat, $C_V$, as we decrease the film thickness $N_z$. \Fig{fig:tcvsnz} 
shows how $T_c$, extracted from the peak in the 
specific heat, changes as we vary $N_z$ for various values of $t_{yz-yz}$. It can be seen that for a wide range of values of this hopping parameter, 
we are able to completely suppress $T_c$ for bilayer samples. We thus see that as we decrease the film thickness, we will tune the system through a 2D Ising quantum critical point. This provides
another promising avenue to suppress $T_c$ and to look for
signatures of octupolar quantum criticality.

\section{Discussion}\label{sec:discuss}
We have discussed a simple tight-binding model with spin-orbit coupling and interactions 
which leads to ferro-octupolar order in the ordered double 
perovskite osmates 
Ba$_2$ZnOsO$_6$, Ba$_2$CaOsO$_6$, and Ba$_2$MgOsO$_6$. Our calculations, which use exact diagonalization and an exact Schrieffer-Wolff
transformation and MC simulations, show that we can capture the octupolar order and its high transition temperature as
observed in experiments. In addition, we have shown how strain and thin film geometries can induce transverse fields which
suppress the octupolar ordering 
temperature, potentially revealing 
an Ising quantum critical point. Our work has implications for a broad class of materials, including $d$-orbital transition metal 
oxides and $f$-orbital heavy fermion systems, where such multipolar orders may be accessible. For instance, NpO$_2$ is a well-known example
of an fcc lattice material which hosts higher-rank multipolar order
with time-reversal symmetry breaking \cite{MultipolarRMP2009,HauleKotliar2009,SantiniNpO2_PRL2000,NpO2TripleQ_PRL2002,Fazekas_PRB2003, NpO2NMR_PRL2006}. 
Octupolar order of the $T_{xyz}$ type explored here
has also been proposed in the Pr(Ti,V,Ir)$_2$(Al,Zn)$_{20}$ compounds where Pr moments live on the diamond lattice \cite{SBLee2018,Patri2019}. In future
work, it would be important to extend our work to understand the origins of multipolar order in these materials. Indeed, there
may be a large set of lattice geometries where such physics
of multipolar order in $d^2$ transition metal compounds
would be worth exploring, as highlighted by Khaliullin, 
{\it et al} \cite{Kee2021}.
More broadly, our work shows that understanding
the impact of thin film geometries, surfaces, and interfaces, in promoting or suppressing multipolar orders in non-Kramers doublet systems may be a fruitful research direction.

{\it Note added:} During completion of this manuscript, we became aware of the recent preprint by Churchill and Kee  
\cite{churchill2021quadrupolar} which has partial overlap with our work. Mainly, these authors identify the combination 
of two different intra-orbital hoppings as favoring octupolar order, as we also independently discovered in our 
work. However, in significant contrast to our work which finds
wide regimes of octupolar order, they find 
evidence for dominant quadrupolar orders.
These quadrupolar orders do not naturally account
for the broken time-reversal symmetry observed in the osmate double perovskites.

\section{Acknowledgments}
We thank F. Lasse Buessen for help with the MC simulations. We thank Giniyat Khaliullin and Hae-Young Kee for useful
discussions.
This work was supported by the Natural Sciences and Engineering Research Council of Canada. Exact diagonalization computations were performed 
on the Niagara supercomputer at the SciNet HPC Consortium. SciNet is funded by: the Canada Foundation for Innovation; the Government of Ontario; Ontario Research Fund - Research Excellence; and the University of Toronto. MC simulations were conducted on the Cedar supercomputer, 
one of Compute Canada's National systems, located in Simon Fraser University.

\appendix 

\section{Choosing $\PHo$ basis states in the Schrieffer-Wolff (SW) process} \label{sw_appendix}

An important step in the SW process is ensuring that we use basis states for $\PHo$ which can be interpreted as direct products of (pseudo)spin states on the two sites. This enables us to interpret the extracted pseudospin Hamiltonian as an direct-product type interaction between the multipole moments in the doublet manifold. While there are many ways to do this, the simplest is to leverage the fact that the $\PHo$ Hamiltonian is decoupled, meaning that the basis states can be constructed out of the single site states. 

At a single site level, we can ensure that the ``up" and ``down" pseudospin states from \Sec{singlesite} correspond to eigenstates of $\tau_z$ ($\propto 3 J_z^2-J^2$), by adding a term to the Hamiltonian which couples an infinitesimal magnetic field to this operator. This weakly breaks the two-fold degeneracy and the basis states of $\PHo$ can simply be built using the direct product of the aforementioned states. 

\section{Symmetry considerations for the Pseudospin Hamiltonian} \label{symm_appendix}
As stated in the main text, the following correspondence exists between $J=2$ angular momentum multipole moments and the Pauli matrices in the low energy non-Kramers doublet manifold: \begin{align}
\begin{split}
{1 \over 2\sqrt{3} }(J_x^2-J_y^2) &\rightarrow \tau_x \\
-{1 \over 3 }\overline{J_x J_y J_z} &\rightarrow \tau_y \\
-{1 \over 6}{(3 J_z^2-J^2)} &\rightarrow \tau_z 
\end{split}\ .
\end{align}
Therefore, the most general spin-1/2 Hamiltonian takes the form (sites are numbered 1 and 2):
\bea 
H_{\rm spin} = {1 \over 4} \mathcal{K}_{ab} \left( \tau_a \otimes \tau_b \right). 
\eea 
For a bond in the $xy$ plane, the following symmetry considerations heavily constrain the form of $\mathcal{K}$:
\begin{enumerate}
    \item Inversion symmetry about the center of the bond exchanges the site indices, implying that $\mathcal{K}$ is symmetric, i.e. 
    \bea
    \begin{split}
    \mathcal{K}_{xy} = \mathcal{K}_{yz} \\
    \mathcal{K}_{yz} = \mathcal{K}_{zy} \\
    \mathcal{K}_{zx} = \mathcal{K}_{xz} 
    \end{split}\ .
    \eea
    \item Under the $M_z$ mirror transformation (the mirror plane is $z=0$), we have 
   \bea 
    \begin{rcases}
        J_z \rightarrow -J_z \\
        J_x \rightarrow -J_y \\
        J_y \rightarrow -J_x 
    \end{rcases}
    \implies 
    \begin{cases}
        \tau_x \rightarrow -\tau_x \\
        \tau_y \rightarrow -\tau_y \\
        \tau_z \rightarrow \tau_z 
    \end{cases}\ .
    \eea
    For the above to be a symmetry of the system, we must have that $\mathcal{K}_{zx} = \mathcal{K}_{yz} = 0$
    
    \item The system is time-reversal symmetric. Under the time reversal operation we have the following transformations:
    \bea
    \begin{split}
    \tau_y \rightarrow -\tau_y \\
    \tau_x,\tau_z \rightarrow \tau_x,\tau_z
    \end{split}\ .
    \eea
    This implies that $\mathcal{K}_{xy} = \mathcal{K}_{yz} = 0$
\end{enumerate}
The above points, when considered together, lead to the vanishing of all the off-diagonal elements in $\mathcal{K}$.
This leaves us with an $XYZ$ Hamiltonian of the form 
\bea 
\mathcal{K} = 
\begin{pmatrix}
\mathcal{K}_{xx} & 0 & 0 \\
0 & \mathcal{K}_{yy} & 0 \\
0 & 0 & \mathcal{K}_{zz}
\end{pmatrix}\ .
\eea 
In the main text, we have renamed $\mathcal{K}_{xx} \rightarrow \mathcal{K}_{Qx}$, $\mathcal{K}_{yy} \rightarrow \mathcal{K}_{O}$, and $\mathcal{K}_{zz} \rightarrow \mathcal{K}_{Qz}$. The exchange
Hamiltonian in the $zx$ and $yz$ planes can be conveniently
obtained using $C_3$ rotations about the cubic (111) direction.

\section{Intersite Coulomb matrix elements}\label{app:Uab}

\textcolor{blue}{
To evaluate matrix elements for the intersite direct 
Coulomb interaction, we must compute integrals of the form
\begin{equation}\label{eq:Vab}
\mathcal{V}_{ab} = {e^2 \over 4\pi\kappa\epsilon_0} \int d\mathbf{r} \int d\mathbf{r'} \text{ } {|\psi_1^a(\mathbf{r})|^2 \text{ } |\psi_2^b(\mathbf{r'})|^2 \over |\mathbf{r} - \mathbf{r'}|},
\end{equation}
where $\kappa$ is the dielectric constant (we set $\kappa=10$), and $\psi_1^a$ ($\psi_2^b$) refers to the wavefunction of an electron in orbital $a$($b$) centered on site 1(2). We use Hydrogen-type $d$ orbital wavefunctions, of the form $\psi^a(\mathbf{r}) = R(r) T^a(\theta,\phi) $, where $R(r)$ refers to the radial part of the wavefunction, and $T^a$ is the tesseral harmonic associated with the $a$ orbital. The detailed form of these functions for the osmate
double perovskites are given in Appendix C of Ref. \cite{voleti2020}.
}

To compute ${\cal V}_{ab}$ in \eq{eq:Vab}, we use a MC integration method. We interpret the integral as a random-walk in a 6-dimensional (6D) space; each point $\vec{x}$ in the 6D space is defined as 
$\vec{x} = (\vec{r}_x, \vec{r}_y, \vec{r}_z, \vec{r}'_x, \vec{r}'_y, \vec{r}'_z)$
in terms of the original position vectors appearing in \eq{eq:Vab}. The random-walk is performed obeying the rules of a Markov process guided by the joined probability distribution 
\begin{align}
    p_{ab}(\vec{x}=(\vec{r},\vec{r'})) = |\psi_1^a(\mathbf{r})|^2 |\psi_2^b(\mathbf{r'})|^2.
\end{align}
Starting from a randomly chosen $\vec{x}_{n=0}$, a sequence of points $\{\vec{x}_0,\cdots,\vec{x}_n\}$ is generated by taking steps $\vec{dx}$ of a sufficiently small length in a randomly-chosen direction, so that upon acceptance of the step we get $\vec{x_{n+1}} = \vec{x}_n + \vec{dx}$. A proposed step $\vec{dx}$ is directly accepted if $p_{ab}(\vec{x}_n+\vec{dx}) / p_{ab}(\vec{x}_n) \geq 1$. If $p_{ab}(\vec{x}_n+\vec{dx}) / p_{ab}(\vec{x}_n) < 1$, a random number $r$ is picked from the uniform distribution defined on the interval $[0,1]$. The step $\vec{dx}$ is accepted if the number $r\leq p_{ab}(\vec{x}_n+\vec{dx}) / p_{ab}(\vec{x}_n)$, otherwise the process is repeated for a new step $\vec{dx'}$. We allow the random-walk to proceed for a large number ($O(10^6)$) of MC steps. The value of the integral in \eq{eq:Vab} is then estimated using
\begin{align}\label{eq:Vab_MC}
    \mathcal{V}_{ab} \approx
    {e^2 \over 4\pi\kappa\epsilon_0} \frac{1}{N_\text{steps}-N_{\text{skip}}}\sum_{n=N_{\text{skip}}}^{N_\text{steps}} \frac{1}{|\vec{r}_n -\vec{r}'_n|},
\end{align}
where $\vec{r}_n$ ($\vec{r}'_n$) is the obtained from the first (last) three components of $\vec{x}_n$ and $N_\text{steps}$ is the number of MC steps. We also skip the first $N_{\text{skip}}$ steps to neglect transient contributions due to the choice of the initial point $\vec{x}_0$. We used $N_{\text{skip}}\sim O(10^3)$ for our calculations. The total number of $N_\text{steps}$ were determined by monitoring when the fluctuations of $\mathcal{V}_{ab}$ in \eq{eq:Vab_MC}, resulting from the random-walk, reduced to less than $1\%$ of the average value.

For a pair of nearest-neighbor sites in the $xy$-plane, and
working in the basis
$\{ yz , zx , xy , x^2\!-\!y^2 , 3z^2\!-\!1 \}$, we obtain 
the Coulomb interaction matrix(in eV):
\begin{equation}
\mathcal{V} = 
\begin{pmatrix}
  0.231 && 0.237 && 0.244 && 0.239 && 0.233 \\
  0.237 && 0.231 && 0.244 && 0.239 && 0.233 \\
  0.244 && 0.244 && 0.261 && 0.260 && 0.245 \\
  0.239 && 0.239 && 0.260 && 0.254 && 0.239 \\
  0.233 && 0.233 && 0.245 && 0.239 && 0.232 
\end{pmatrix}.
\end{equation}
We note that this matrix has an orbital independent value
$\sim 0.25$\,eV, and orbital-{\it dependent} parts on 
the scale of $\sim 30$\,meV.
Adding this Coulomb term to the two-site Hamiltonian as $\sum_{\alpha\beta} {\mathcal V}_{\alpha\beta} n_{1,\alpha} n_{2,\beta}$, where $n_{i\alpha}$ 
is the total electron number in orbital $\alpha$ at site $i$,
we use ED and the Schrieffer-Wolff method to recompute
the exchange interactions and effective fields.
We find that the two-site exchange interactions are nearly
unaffected since the scale of ${\mathcal V}_{\alpha\beta}$ 
is much smaller
than the on-site Hubbard interaction $U=2.5$\,eV. However,
the orbital-{\it dependent} part of
${\mathcal V}_{\alpha\beta}$
results in an extra effective field $\propto \tau_z$
on the non-Kramers
doublet. To see this, we note that mean field theory
yields an on-site term at site $i=1$ given by 
$\sum_{\alpha\beta} {\mathcal V}_{\alpha\beta} 
n_{1,\alpha} \langle n_{2\beta} \rangle$. Upon projection
to the non-Kramers doublet, this acts as an extra effective 
field. Our computations show that the scale of this field 
is $\sim 8$\,meV,
which is larger than the field produced by two-site
exchange interactions alone. Thus, while these fields cancel
in the cubic bulk, the Coulomb-enhanced 
surface fields are
strong enough to overcome the ferro-octupolar exchange
and kill the octupolar order for
sufficiently thin films as shown in the paper.

\bibliography{octupolar}

\end{document}